# MICROSTRUCTURE AND STRUCTURAL DEFECTS IN MgB$_2$ SUPERCONDUCTOR

Y. Zhu, L.Wu, V.Volkov, Q.Li, G.Gu*, A.R. Moodenbaugh, M.Malac, M. Suenaga, and J. Tranquada*

Materials Science Division, Brookhaven National Laboratory, Upton, NY 11973

* Physics Department, Brookhaven National Laboratory, Upton, NY 11973

(May 12, 2001)

## ABSTRACT

We report a detailed study of the microstructure and defects in sintered polycrystalline MgB$_2$. Both TEM and x-ray data reveal that MgO is the major second-phase in our bulk samples. Although MgB$_2$ and MgO have different crystal symmetries, being P6/mmm and Fm-3m, respectively, their stacking sequence of Mg and B (or O) and lattice spacings in certain crystallographic orientations are very similar. The size of MgO varies from 10~500nm, and its mismatch with the MgB$_2$ matrix can be a source for dislocations. Dislocations in MgB$_2$ often have a Burgers vector of <100>. 1/3<1-10> and 1/3<210> partial dislocations and their associated stacking faults were also observed. Since both dislocations and stacking faults are located in the (001) basal plane, flux pinning anisotropy is expected. Diffuse scattering analysis suggests that the correlation length along the c-axis for defect-free basal planes is about 50nm. (001) twist grain-boundaries, formed by rotations along the c-axis, are major grain boundaries in MgB$_2$ as a result of the out-of-plane weak bonding between Mg and B atoms. An excess of Mg was observed in some grain boundaries. High-resolution nano-probe EELS reveals that there is a difference in near edge structure of the boron K-edge acquired from grain boundaries and grain interiors. The change at the edge threshold may be suggestive of variation of the hole concentration that would significantly alter boundary superconductivity.

## 1. Introduction

The recent discovery of the 39K superconductivity in the binary intermetallic compound, magnesium diboride (MgB$_2$), by Akimitsu and co-workers [1] evoked a new enthusiasm for understanding the mechanism of superconductivity and the structure-property behavior of this new class of materials. MgB$_2$ possesses the highest superconducting transition temperature T$_c$ outside the cuprate- and C$_{60}$-related superconductors, and its superconductivity appears to follow the BCS theory, i.e., mediated via electron-phonon coupling. It was reported that the coherence length in MgB$_2$ is longer than that in the cuprates [2]. Also, unlike curpates, grain boundaries are strongly coupled and current density is determined by flux pinning [2,3]. Despite the phase inhomogeneity, porosity, untextured, nanoscale subdivisions of the not-yet-optimized MgB$_2$ ceramics, magnetization and transport measurements show that the polycrystalline samples can carry large current densities circulating over lengths of many grain sizes [3,4]. More recently, very high value of critical current densities as well as the critical fields were observed in thin films [5,6]. These properties hold tremendous promise for a wide range of large-scale and electronic applications, and suggest that the underlying microstructure can be intriguing. Immediately following the discovery, dozens of papers were published [1-13]; however, most of which dealing with synthesis, measurements of physical properties, and theory. There were few studies of microstructure and structural defects [11], which are extremely important for understanding the superconducting behavior. In this article, we report such an investigation using transmission electron microscopy (TEM) on sintered

MgB$_2$ pellets. Structural defects, such as second-phase particles, dislocations, stacking faults, and grain boundaries were observed and analyzed using electron diffraction, electron-energy loss spectroscopy (EELS), high-resolution imaging, and structure modeling. Our analysis of the fundamental structural behavior in MgB$_2$ provides detailed insight on the structure-properties relationship of the new superconductor not only for samples sintered under Mg vapor, but also for those fabricated via other synthesis procedures.

## 2. Sample preparation and magnetic properties

Bulk samples were prepared using commercial MgB$_2$, which then was pressed into pellets 6mm in diameter. The pellets were sealed into Al$_2$O$_3$ tubes with pure Mg and sintered under Mg vapor pressure of approximately 300 Torr at 1050 $^o$C for 24 hours, followed by furnace cooling. The sintered pellets are gray and Mg-rich. Several center-cut and polished specimens were examined using an optical microscope, they appeared shiny and very dense (porosity is less than 5%) with average MgB$_2$ grain size of ~ 10 μm. The superconducting properties were studied using a Quantum design MPMS SQUID magnetometer and standard 4 point transport measurements. Fig.1(a) shows the shielding (zero field cool) and Meissner effect (field cool) of the MgB$_2$ specimen measured 2 Oe external field. The specimen has a sharp transition with mid-point $T_c$ = 37.2 K and transition width of 0.7 K. A full bulk shielding (~100%) was observed after taking into account the demagnetization factor of the specimen.

Fig.1(b) shows the magnetic field dependence of critical current density $J_c$ of the specimen at various temperatures. This magnetically determined $J_c$ was obtained by applying the standard Bean critical state model to the measured magnetic hysteresis of the specimen at applied external field up to 5 T, where the average grain size of 10 μm was used. At 5 K and in self-field, the $J_c$ for our specimen reaches as high as 5 x 10$^7$ A/cm$^2$. If the entire measured specimen dimension of ~ 2mm is assumed, the value of $J_c$ is scaled down by a factor of approximately 200. Then, the corresponding value of $J_c$ at 5 K and self-field would be 2.5 x 10$^5$ A/cm$^2$, which is comparable to that reported by other groups in the literature [2,3,8].

A standard θ~2θ x-ray powder diffraction technique was used to identify the bulk phase, composition and lattice parameter. X-ray diffraction patterns were collected with Cu Kα radiation (λ=0.154178nm) over a 2θ range from 10° to 100° at a step width of 0.02°, after a diffracted beam monochromator. Detailed microstructure and structural defects were characterized using a JEOL3000F electron microscope with a field-emission source operated at 300kV. After mechanical polishing down to a thickness of 15μm, TEM samples were prepared via a standard ion-milling procedure at 3.5kV and 10° glancing angle at liquid N$_2$ temperature. The microscope has an image resolution of 0.165nm and energy resolution of 0.7eV. The data acquisition system is equipped with a TV-rate camera, a slow scan CCD and Fuji Imaging Plates. The CCD camera is integral to the Gatan energy filter and energy-loss spectrometer. For chemical analysis, we used both energy-dispersive x-ray spectroscopy (EDX) and electron energy-loss spectroscopy (EELS). Since both boron and magnesium (Z=5 and 12, respectively) are light elements, EELS has much higher sensitivity to probe them than EDX does. Furthermore, the absorption of electron-induced soft x-rays from light elements poses a significant limitation on quantitative EDX analysis. Thus, we relied heavily on EELS for our chemical analyses. Steps were taken to minimize contamination which might affect nanoprobe characterization. The samples were examined either at liquid N$_2$ temperature or, if at room-temperature, underwent a plasma cleaning process for 20-30 minutes using argon and oxygen gases which significantly reduced carbon contamination.

## 3. Microstructure and structural defects
### 3.1 Microstructure

Fig.2 shows the x-ray diffraction spectrum of the powder sample ground from a sintered pellet. All the reflections can be unambiguously indexed to three phases, with roughly 70% of MgB$_2$ and 30% of periclase (MgO) and metal magnesium. No other phases, such as MgB$_4$ and MgB$_7$, were detected. The lattice parameters for MgB$_2$ were determined to be a=0.3086(5)nm and c=0.3518(5)nm, consistent with ref [1]. Metal magnesium metal was not observed in the TEM samples; it was probably oxidized during the ion-milling and plasma-cleaning processes. TEM reveals that MgB$_2$ has space group P6/mmm, possessing a simple hexagonal structure common among the borides [14], being consistent with our x-ray diffraction observations. This crystal structure consists of honeycomb-net planes of boron, separated by triangular planes of Mg, with the center of a boron honeycomb lying both directly above and below each Mg atom. In Fig.2 and in the following analysis, the three-index notation, *hkl*, is used to index the $MgB_2$ crystal. Although the four-index notation, *hkil*, brings in the six-fold crystal symmetry, it does not directly represent the reciprocal relation between diffraction and image which is particularly important for TEM. In the three-index system, a reciprocal vector is always perpendicular to its corresponding lattice plane, but not necessary parallel to the direction vector of the same index in real space.

Generally, the individual $MgB_2$ crystals (or grains) of our samples seemed rather clean and dense, as suggested by their sharp Kikuchi lines shown in convergent beam diffraction. However, because of the existence of the aggregates of solidified MgO and many poorly sintered, and pooly connected $MgB_2$ grains (Fig.3a), overall the samples were mechanically weak. Some parts appeared porous and crumbly, similar to the sintered, oxygen-deficient, bulk samples of $YBa_2Cu_3O_{7-\delta}$ ($\delta>0.5$) as assessed by mechanical polishing and optical microscopy. We take advantage of the small probe of TEM and focus on stoichiometric $MgB_2$ grains that act like single crystals under our microscope. The majority of the $MgB_2$ grains we observed are faceted along the basal plane, and many of them are shaped like platelets with an aspect ratio of about 1:3 (Fig.3a), suggesting a two-dimensional character consistent with the layered nature of the structure. The long axis of the platelets is about 2μm on average. For grains without such a shape, the averaged size is about 1.5μm. Few crystal grains are free of defects. The major defects in $MgB_2$ crystals are second-phase particles, dislocations, including stacking faults associated with partial dislocations, and grain boundaries. We discuss the observations and analysis of these defects below.

### 3.2 Second-phase particles

The major second phase in our samples is MgO, although particles containing other elements, such as Fe, were also observed. The impurity atoms likely come from the original commercial powder. The size of the MgO particles had a wide distribution, ranging from 10-500nm (Fig.3a-c). The large ones often assemble the $MgB_2$ grain, although MgO grains can be faceted along the {220} and {422} planes (Fig.3b). Both x-ray and TEM diffraction show that MgO has fcc structure belonging to space group of Fm-3m. Despite the different crystal symmetries of MgO and $MgB_2$, cubic and hexagonal, respectively, their diffraction patterns in some orientations look identical (Fig.4a,b), so that the same crystal must be tilted to different zone axes and its symmetry then examined to unambiguously distinguish between MgO and $MgB_2$. For example, even with the presence of the unique dumbbell configuration for $MgB_2$ in the hexagonal structure, the lattice symmetry and in-plane spacings projected along the [1-1-1] and [2-21] directions are very similar to those of MgO projected along the [001] and [-111] directions, respectively (Fig.4). Fig.4a shows a calculated diffraction pattern superimposing $MgB_2$ (1-1-1)* on MgO (001)*. We see that all of the low-order reflections are coincident. The notable difference is the intercept angle of the "square" lattices (as marked in Fig.4a,b). The angle for MgO projected along the [001] is 90°, while for $MgB_2$ along the [1-1-1] direction is 90°±α, where α= 2.8°. A similar situation exists for the projected lattice plane for $[-111]_{MgO}$ and $[2-21]_{MgB2}$. The former has a true 3-fold symmetry, while the latter has an the intercept angle 60°±β with β=3.1°. In both cases, the difference in their corresponding lattice spacings is negligibly small. For example, $d_{220(MgO)}$=0.1489nm, $d_{1-12(MgB2)}$=0.1470nm; $d_{224(MgO)}$=0.0860nm and $d_{-114(MgB2)}$=0.0836nm. Such orientation relationships may imply the possible low-energy epitaxial growth of the two phases and the existence of good lattice-match at the MgO and $MgB_2$ interfaces, as seen in Fig.3c where the interfacial misfit dislocations and the associated strain are not visible. Care has to be taken to identify them. For small particles where systematic tilt is not feasible, EDX and EELS can be used to distinguish MgO from $MgB_2$. As seen in Fig.5, besides the pronounced K-edge of boron at 188eV and of oxygen at 532eV for $MgB_2$ and MgO, respectively, the Mg core-loss fine structure is also different for the two, suggesting the difference in bonding characteristics of magnesium with boron and oxygen. The insets of Fig.5 show the Mg K-edges after background subtraction. For clarity, the core-loss intensity of the Mg edge in Fig5b was amplified by a factor of 8.5.

### 3.3 Dislocations and stacking faults

Dislocations are the predominant defects in $MgB_2$. Most of them are perfect dislocations with mixed edge and screw components lying in the a-b basal planes normal to the c-axis. They sometimes interact with second phase particles. Fig.6(a-d) shows diffraction contrast of the same area containing straight dislocations, dislocation loops and MgO particles. The images were recorded from the same area near the [10-1] direction, i.e., 41° off the [001] direction, with different operating reflections. Characterization using the **g**×**b** extinction rule suggests that most are screw dislocations with a Burgers vector of [100], and as evident when **g**=010 was used, they are out-of-contrast (Fig.6d). The particles marked by black arrows in Fig.6a are MgO. Thickness fringes observed within the particles suggest they are spherical in shape. Dislocation loops and partial dislocations, both bonding stacking fault and lying in the basal plane, are also present, as marked by small arrows in Fig.6b. The displacement vector of the stacking fault associated with the dislocation loops apparently differs from the Burgers vector of the dislocations that bond them, based on the diffraction contrast under different imaging conditions. Fig.7 are basal-plane dark-field (DF) images (viewed along the c-axis), showing the interaction of dislocations and two small MgO particles, each about 24nm in diameter. The inset of Fig.7(a) is the DF image of one of the particles recorded using the MgO 400 reflection. Tilting experiments reveal that these are general perfect dislocations gliding in the basal plane. When the 030 reflection was used, the dislocations are out-of-contrast, suggesting a [100] Burgers vector. Diffraction analysis reveals an orientation relationship between the two phases with $(111)_{MgO}//(101)_{MgB2}$, resulting an in-plane lattice mismatch of $\delta=14\%$. The origin of the dislocation nucleation and propagation associated with the imbedded MgO particles can be attributed to the interfacial strain cause by different thermal contraction rates of MgO and $MgB_2$ and the stiffness of the a-b lattice in $MgB_2$. This is consistent with Jorgensen et al's observations that $MgB_2$ has unusually large anisotropies with larger thermal expansion and compressibility along the c-axis, but not in the basal plane [9].

Figure 8 shows dislocation images viewed along a direction orthogonal from those shown in Fig.7. Here the a-b planes are viewed nearly edge-on. The line contrasts in Fig.8 are attributed to dislocations lying in the a-b plane. When **g**=002 was used, the line contrast disappeared, suggestive of an in-plane Burgers vector. However, a detailed analysis shows that these are not screw dislocations, but are general dislocations in the a-b plane, running from the top to the bottom surfaces of the sample, as evident by the intensity oscillation that is associated with the thickness extinction along the dislocations. In the [010] zone axis (Fig8a), the line contrast represents the projected dislocation contrast in the basal plane. In the area of Fig.8b, marked by an arrow, we see the dissociation of a perfect dislocation into two partials. We note that the double line-contrast cannot be attributed to dynamic diffraction because the split is visible (Fig.8c) even when the imaging conditions **g**×**b**<2 were used [15]. Apparently the dissociation is due to the lattice shift in the basal plane.

In Fig.9, we propose structural models for perfect dislocations and partial dislocations involving both Mg and B sublattices in the system. Fig.9a represents a model for an edge dislocation with a Burgers vector of [100] lying in the basal plane as observed in the experiments. In this case, the lattice shift induced by a dislocation is over two sublattices spacings, i.e., either Mg→Mg, or B→B. For clarity, Fig.9a shows the projected a-c plane and a-b plane of the $MgB_2$, where the small dots represent B atoms, while the big gray and black dots represent Mg atoms at different depth in the crystal. Based on this model, the presence of the edge dislocation can be imagined as the insertion of two Mg-B half-planes with a total spacing of a=0.308nm numerically equal to the modulus of Burgers vector of **b**=[100]. Fig.9b illustrates the dissociation of a perfect **b**=[100] edge dislocation into two Shockley partial dislocations with Burger vectors $\mathbf{b}_1=1/3[1-10]$ and $\mathbf{b}_2=1/3[210]$, bounding a stacking fault in between. Here, we assign A, B, C as magnesium sites and α, β, γ as boron sites, as depicted in Fig.9c. The diagrams (Fig.9c-e) explain how the partials with Burgers vector $\mathbf{b}_1$, $\mathbf{b}_2$, or $\mathbf{b}_3$ change the stacking sequence along the c-axis in a hexagonal crystal. Using this notation the formation of planar fault caused by the dislocation reaction $\mathbf{b}=[100]=\mathbf{b}_1+\mathbf{b}_2= 1/3[1-10]+1/3[210]$ can be expressed as:

…A(βγ)A(βγ)A(βγ) | A(βγ)A(βγ)A(βγ) …
      $b_1$: | B(γα)B(γα)B(γα)B(γα)B(γα) …
          $b_2$: | A(βγ)A(βγ)A(βγ) …

That is, after the dissociation, the stacking sequence shifts back to its original. This is similar to Shockley partials in simple fcc structures. The energy associated with the dissociation can be expressed as $E_b=E_{b1}+E_{b2}+E_{SF}$. Since $E_{[100]}>E_{1/3[1-10]}+E_{1/3[210]}$, it implies that the reaction is energetically favorable, and the energy difference is balanced by the energy of the stacking fault $E_{SF}$ ($E_{SF}= w \times e_{SF}$, where $w$ and $e_{SF}$ are the width and energy per unit length, respectively, of the stacking fault). Since experimentally, only narrowly spaced partial pairs (Fig.6b and Fig.8b,c) were observed, a large value of $e_{SF}$ is suggested. The possible dissociation for other equivalent perfect dislocations can be derived using a general diagram shown in Fig.9e.

The motion of the partial dislocations and the associated stacking fault we described above is conservative. However, a non-conservative fault can also be formed if there is a missing a-b plane of boron or an intercalation of an a-b plane of magnesium at the fault. The situation for the lattice shift is the same for the stacking fault described in Fig.9, except it forms a local magnesium structure. Magnesium is a hcp crystal with the lattice parameter a=0.32nm, very similar to the lattice parameter a=0.31nm for $MgB_2$. The stacking fault associated with the dislocation loop we observed (Fig.6) belongs to this type. The formation of such loops (Frank dislocation loops) can be attributed to the excess Mg in our samples.

The distribution of the faulted basal planes due to the dislocations and stacking faults is best seen when the a-b planes are viewed edge-on, the grains B and C in Fig.10 serving as examples. Electron diffraction, using parallel illumination of a large area similar to grain B, recorded on an image plate, revealed a broadening of the Bragg spots along a direction perpendicular to the basal-plane (Fig.10d). A line scan of the intensity of 300 reflection is shown Fig.10e. The reflection intensity can be fitted into two profiles; one is a mixture of Gaussian and Lorentzian distribution attributed to the Bragg peak, the other is a Gaussian distribution associated with diffuse scattering. The width of the diffuse scattering peak gives an average of 50nm correlated length, i.e, the averaged distance from one non-faulted basal plane to the next.

### 3.4 Grain boundaries

There are two types of grain boundaries in our samples. One represents poorly coupled grains with amorphous material at the boundary, while the other type involves well-coupled grains with structurally intact boundaries, marked as "1" and "2" in Fig.10a, respectively. The amorphous boundary in Fig.10a is a tilt boundary with the basal plane of one of the constituent grains as its boundary plane. The two grains across the boundary are rotated 45° about the [010] axis, as is evident from the line contrast due to the faulted a-b planes (Fig.10a,b). Although the boundary is amorphous overall, local crystalline areas are also visible, often having an epitaxial relationship with the crystal grain, as shown in Fig.10c (the dashed vertical line indicates the interface between left grain and the thick amorphous boundary). As discussed later, these crystalline areas are likely to be MgO. The thickness of the amorphous layer at such boundaries ranges from 10~50nm.

For structurally intact grain boundaries, we observed many (001) twist boundaries, both large-angles and small-angles, with an in-plane rotation about the c-axis across the interface, similar to the grain boundaries in the $Bi_2Sr_2CaCu_2O_8$ superconductor. This can be attributed to the layered nature of the $MgB_2$ structure, with graphite-like boron layers which are separated by hexagonal close-packed layers of Mg. It is consistent with the prediction that the B-B bonds in the basal plane are much stronger than the Mg-B bonds that connect layers of Mg and B atoms [9]. The lattice distortion associated with such boundaries often extends only a few atomic layers and their interfacial structure can be retrieved at different length scales using diffraction contrast, as well as by high-resolution imaging. An example of a 4.1° small-angle (001) grain boundary is shown in Fig.11. Fig.11a displays a two-beam image of the boundary showing the diffraction contrast of the quasi-periodic dislocation array at the interface. Fig.11b is a high-resolution (002) lattice image showing alternating Mg-B layers running across the boundary. As characterized by Kikuchi patterns, the major components of the boundaries are the 3.7° twist component about the [001] axis and a small tilt component 0.8° about the [120] axis. A careful examination of the Kikuchi pattern reveals another tilt component of 1.4° about the [100] axis which is accommodated by inserting extra a-b planes (marked as an edge dislocation in Fig.11c) at the boundary. Based on the Frank formula, $d = sin\bm{q} / b$, where b is the amplitude of Burgers vector of the edge dislocations, $\bm{q}$ the tilt angle, and $d$ the dislocation spacing, the calculated dislocation spacing (14nm) and Burgers vector ([001]) are consistent with our experimental observations. The other components of the boundary may be accommodated by two sets of interfacial screw dislocations, which are not visible in the images (Fig.11a-c) when they are not viewed edge-on. In Fig.11c, the [100] direction of the left crystal was tilted parallel to the incident beam, while the right crystal was off the [100] axis. A high-resolution image of the boxed area in Fig11c is shown in Fig.11d, where the Mg atoms are viewed as bright dots and B atoms weak white dots. The embedded image is a simulated one with four unit-cells in the projection, showing good agreement with the experiment. The projected boron dumbbell with a spacing of 0.089nm is not resolved in the image.

Quantitative nano-probe EELS was carried out on several grain boundaries and compared with the results from grain interiors. To achieve a small probe with sufficient beam current and energy resolution, the microscope was set in the nano-beam-diffraction (NBD) mode with a nominal probe size of 0.6nm (the actual FWHM of the probe is about 1nm), and a collection angle of 10mrad. To improve the signal/noise ratio, the spectra were acquired in diffraction mode using an accumulative acquisition method with the maximum intensity in each acquisition being just slightly below the saturation of the CCD camera.

Initial elemental quantification was done by standard power-law background subtraction at the elemental edges, followed by integration of the core-loss intensity to obtain the intensity ratio, such as $I_K^{Mg}/I_K^B$. The formula $N^{Mg}/N^B = (I_K^{Mg}/I_K^B)(s_K^B/s_K^{Mg})$ was used to obtain the areal concentration ratio, $N^{Mg}/N^B$. The K-ionization cross sections were calculated using the hydrogenic approximation for generalized oscillator strength [16]. For $E_0$ =300keV and $b$=10mrad, we obtain $s_K^B/s_K^{Mg} = 86.64$ for an energy window $D$=50eV, and $s_K^B/s_K^{Mg} = 97.98$ for $D$=30eV. Although a larger window is more susceptible to the errors due to the background subtraction, it is less sensitive to the near-edge fine structure. The averaged integrated intensities using both windows were used. Tables I and II list the variation of concentration ratios for the structurally intact boundaries and amorphous boundaries, respectively.

Most of the quantitative measurements were obtained from thin regions with a thickness about $0.5\lambda_m \sim \lambda_m$ (where $\lambda_m$ is the inelastic mean-free-path). Table 1 gives the concentration ratio of $N^{Mg}$ to $N^B$ for three structurally intact boundaries of $MgB_2$ using 1nm-probe EELS, where GB denotes the grain boundary and GI denotes the grain interior about 100nm away from the boundary. Each measurement represents an average of five data points. For GBI and GBII, we observe a ~50% increase in Mg /B ratio at the boundaries, while in GBIII we saw no significant change. Evidently, Mg segregation at the boundaries is sensitive to local structure. The relation to the boundary geometry including misorientation angle and core-structure of interfacial dislocations will be the subject of a further study. However, it is important to note that these non-stoichiometric grain boundaries are similar to those found in $Nb_3Sn$ [17,18] and can be the site of strong flux pinning when the grain sizes are significantly reduced [a].

At amorphous boundaries (Table 2), besides seeing variation of Mg, we observed a significant excess of oxygen. The oxygen in the grain interior is barely detectable, while strong oxygen peaks were observed at the boundaries. The concentration ratio of oxygen to boron $N^O/N^B$ at the two boundaries were determined to be 0.451 and 0.702, more than one order of magnitude higher than at the grain interiors. The high concentration of oxygen may be associated with an oxide wetting-phase lying between poorly sintered $MgB_2$ grains. It may be formed by oxidation of Mg vapor at high temperatures during sintering.

The fine structure of the boron K pre-edge was also studied using high-resolution EELS for large-angle structurally intact grain boundaries. Fig.12 is an example revealing the difference between the fine structure at the grain boundary (GB) and at the grain interior (GI). The sharp boron peak at 202eV is accompanied by a broad peak centered at 223.5eV. The spectra were acquired from relatively thick areas, $1.8\lambda_m$ and $1.6\lambda_m$, respectively, in order to maximize the interfacial volume with respect to the volume of the amorphous layers formed during ion milling at the top and bottom surfaces of the sample. Fortunately, plural scattering from a thick region has little effect on the pre-edge fine structure of an elemental edge. On the other hand, it can significantly modify the core-loss intensities above 210eV, since the first plasmon peak of $MgB_2$ is at 20eV. The dashed lines in the spectra represent the core-loss intensity of the post-edge after the removal of the plural scattering, based on the Fourier-ratio devolution method, i.e., the spectrum Fourier transform is divided by the low-loss Fourier transform, followed by an inverse Fourier transform. It is interesting to note that there are two weak peaks in the boron pre-edge spectra at about 190.5eV and 194.5eV, respectively. The former is seen only in the spectrum acquired from the grain interior, while the latter is more clearly visible from the grain boundary. The change in the boron pre-peak intensity may be related to a change of hole concentration in the area. Like $YBa_2Cu_3O_7$, $MgB_2$ is considered to be a hole-doped superconductor, based on Hall effect measurements [13]. In $YBa_2Cu_3O_7$, the change in pre-peak intensity of the oxygen K-edge at 528 eV and at 530 eV was related to the boundary superconductivity and weak-link behavior [19,20]. The former corresponds to the hole density (1s→2p transition) and attains a maximum for optimum superconductivity, while the later is associated with a hybridized d state (1s→3d transition), and has a maximum intensity in the insulating non-superconducting regime. Recently, An and Pickett [10] calculated the effects of various phonon modes of the electronic structure of $MgB_2$. They concluded that superconductivity results almost exclusively from in-plane boron bands that contribute strongly to the Fermi-level density of states because of the two-dimensional nature of the material. They found that there are non-bonding π band ($p_z$) and bonding σ band ($sp_xp_y$) associated with boron atoms, and that the σ to π charge-transfer results in hole doping. In their calculations the σ band extends to about 1eV above the Fermi level. We believe this is the feature we observe at 190.5eV, which is then associated with variation in the hole concentration. The lack of the pre-peak intensity at the boundary resembles the situation in the $YBa_2Cu_3O_7$ where the disappearance of the pre-edge at the oxygen K-edge is the signature of a grain boundary with degraded superconducting properties. A systematic study in $MgB_2$, ideally using bi-crystals, of electronic structure and superconductivity of grain boundaries, has not yet been conducted. If the boundaries are strongly coupled, as reported in ref.[2,3], it suggests that the width of the interfacial hole-depletion region may be smaller than the coherence length of $MgB_2$, which was estimated to be 5nm [2, 21]. Optimistically, the hole-deficiency might enhance flux pinning at the boundaries. At any rate, it is likely that such hole-density variations will have a profound effect on superconducting transport properties across the boundaries in this binary intermetallic compound.

## 4. Conclusions

The structural defects in sintered $MgB_2$ have been analyzed. The major second phase was found to be MgO particles with a wide size-distribution, ranging from 10-500nm in diameter. Depending on their crystallographic orientation relative to the $MgB_2$ matrix, these MgO particles can generate dislocations. Most dislocations observed were perfect dislocations with a Burger vector **b**=[100] or [010], lying in the basal plane. Stacking faults bounded with a pair of partial dislocations were also observed. Although both second-phase particles and dislocations can act as strong pinning centers, the pinning strength from dislocations and stacking faults are likely to be confined within the a-b planes, i.e., being weak for flux lying in the a-b plane, but strong along the c-axis. Grain boundaries in $MgB_2$ often have this basal plane as their boundary plane, consistent with the weak bond between boron and magnesium atoms. An excess of Mg was observed even in structurally intact grain boundaries. High-resolution EELS revealed that there is a change at the onset of the boron K-edge at the grain boundaries compared with that from the grain interiors, suggesting a change of the electronic structure that was modified by the Mg-B bonding at the interfaces. Since the correct density of states near the Fermi level is required to achieve superconductivity, a reduction in states may significantly alter the transport properties across the boundaries. This phenomenon may be similar to the change of the intensity of pre-peaks at the oxygen K-edge across the grain boundaries in $YBa_2Cu_3O_7$.

**Acknowledgement**

This work is supported by U.S. Department of Energy, Division of Materials, Office of Basic Energy Science, under Contract No. DE-AC02-98CH10886.

Table 1 EELS measurements of boron and magnesium concentration ratio at structurally intact grain boundaries (GB) and grain interiors (GI). The core-loss cross-section ratios $(s_K^B / s_K^{Mg})_{D=30V} = 0.98 \times 10^2$ and $(s_K^B / s_K^{Mg})_{D=50V} = 0.87 \times 10^2$ were calculated for *300kV* electrons at an collection of *10 mrad*.

|        | $I^{Mg} / I^B$ | $N^{Mg} / N^B$ | $(N^{Mg}/N^B)^{GB} / (N^{Mg}/N^B)^{GI}$ |
|--------|----------------|----------------|------------------------------------------|
| GB I   | 3.7x10$^{-2}$  | 3.24           | 1.64                                     |
| GI II  | 2.3x10$^{-2}$  | 1.98           |                                          |
| GB I   | 3.5x10$^{-2}$  | 3.00           | 1.44                                     |
| GI II  | 2.4x10$^{-2}$  | 2.09           |                                          |
| GB III | 2.5x10$^{-2}$  | 2.15           | 1.06                                     |
| GI III | 2.3x10$^{-2}$  | 2.03           |                                          |

Table 2 EELS measurements of boron and oxygen concentration ratio at amorphous grain boundaries (GB) and grain interiors (GI). The core-loss cross-section ratios $(s_K^B / s_K^O)_{D=30} = 9.98$ and $(s_K^B / s_K^O)_{D=50} = 9.16$ were calculated for *300kV* electrons at an collection of *10 mrad*.

|        | $I^O / I^B$    | $N^O / N^B$ |
|--------|----------------|-------------|
| GB I   | 6.7x10$^{-2}$  | 0.702       |
| GB II  | 4.3x10$^{-2}$  | 0.451       |
| GI     | 3.7x10$^{-3}$  | 0.038       |

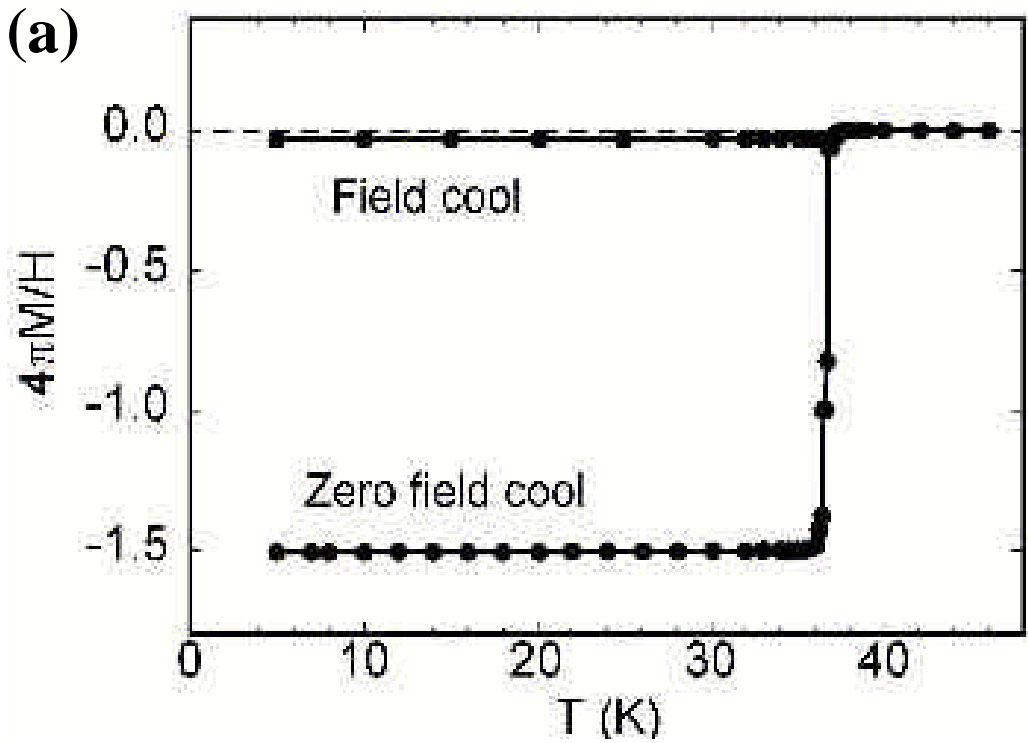

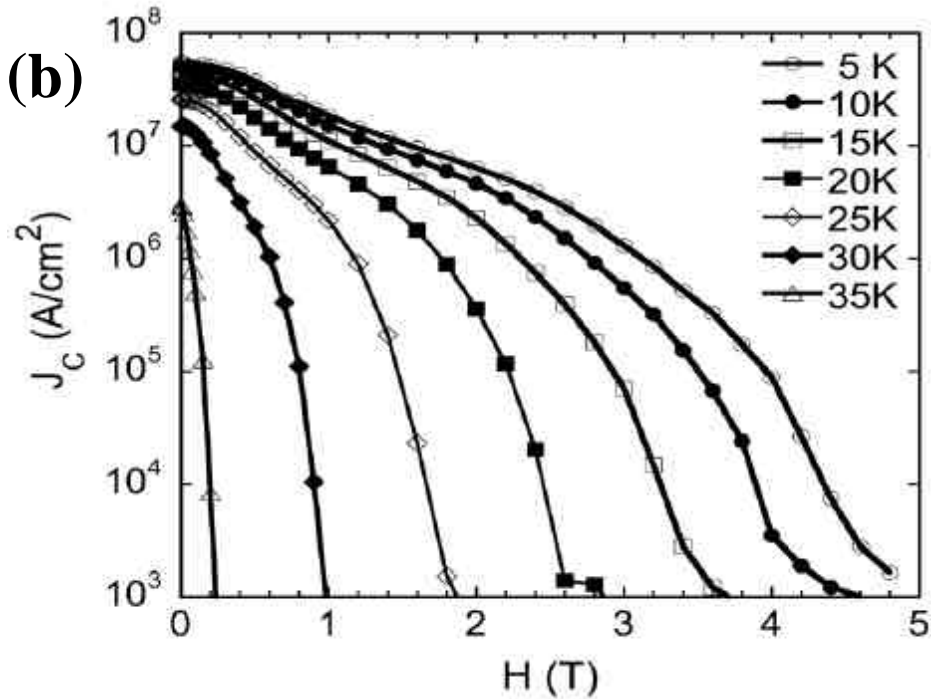

Fig.1  (a) Magnetic shielding (zero field cool) and the Meissner effect (field cool) of the $MgB_2$ sample under external field of 2 Oe.  (b) Magnetic field dependence of critical current density $J_c$ of the $MgB_2$ sample at various temperatures.

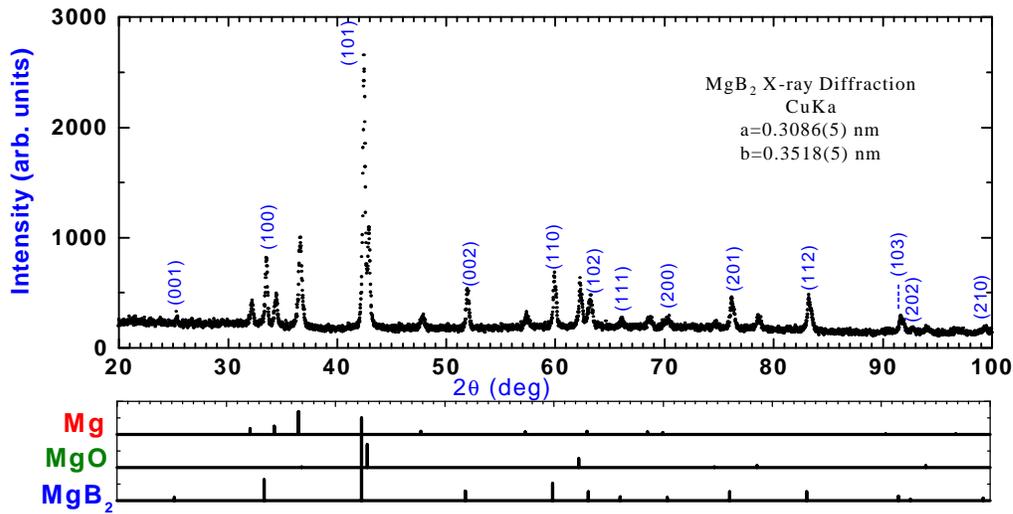

Fig.2 X-ray diffraction measurements showing a mixture of $MgB_2$, MgO, and Mg in the sample. The diffraction peaks are indexed to $MgB_2$, and the peak positions of all three phases are marked below.

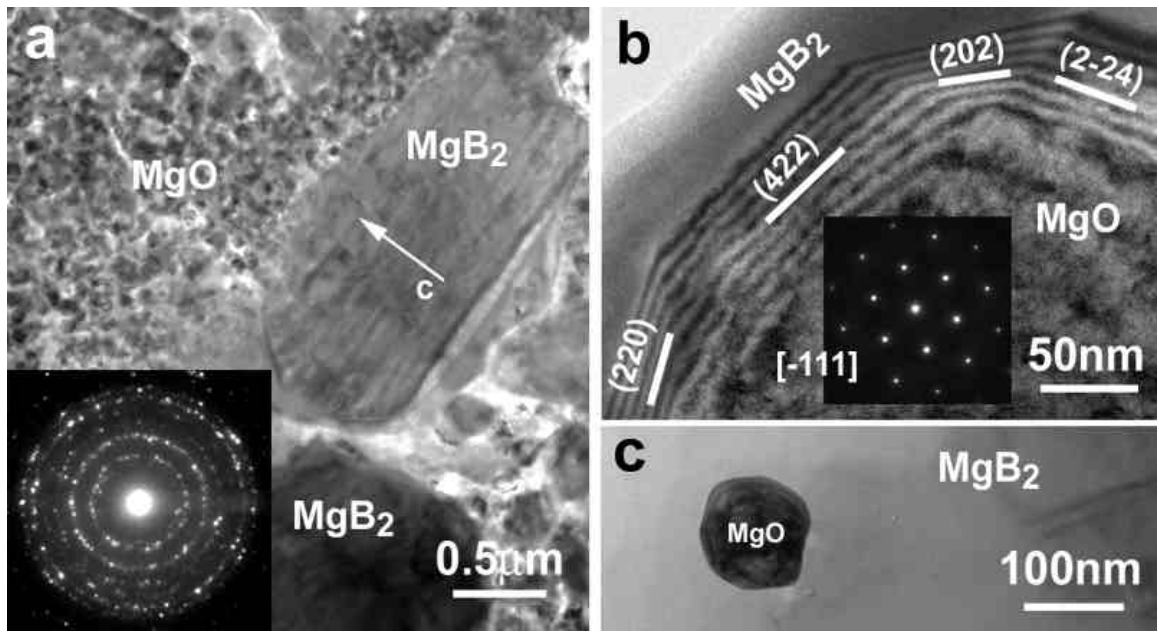

Fig.3 MgO second-phase particles in $MgB_2$. (a) Small aggregated MgO particles solidified from MgO vapor and their corresponding diffraction ring-pattern. The nearby $MgB_2$ platelet exhibits faceting along the basal plane; (b) A large MgO particle viewed along the [-111] three-fold axis showing faceting parallel to the {220} and {422} lattice planes; and (c) An isolated small MgO particle embedded in $MgB_2$ showing little strain contrast at the interphase interface.

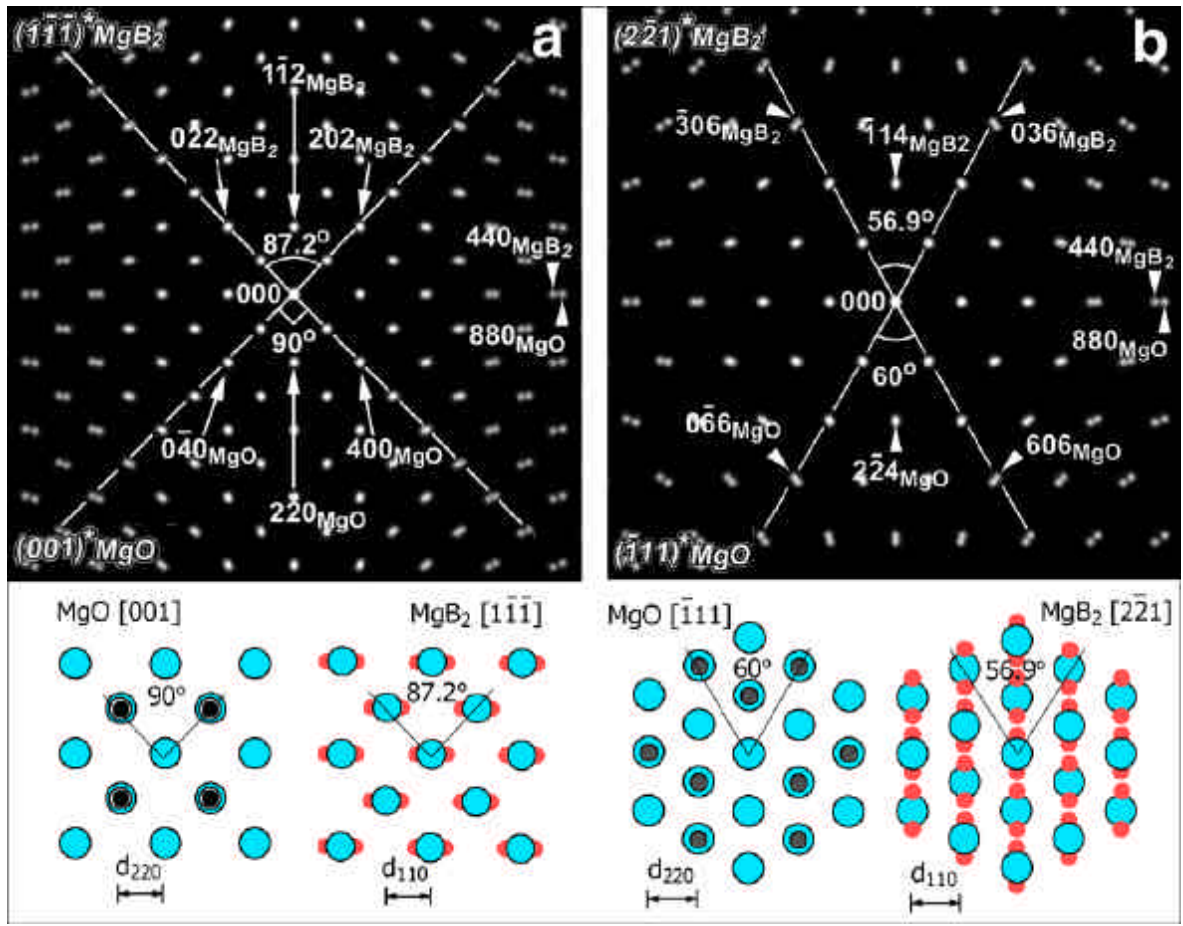

Fig.4 Structural similarity of $MgB_2$ and MgO in two crystallographic orientations. (a) Superimposed diffraction patterns of $(001)^*_{MgO}$ and $(1\text{-}1\text{-}1)^*_{MgB2}$; configurations of Mg and B layers projected along the $[001]_{MgO}$ and $[1\text{-}1\text{-}1]_{MgB2}$ directions are shown below. Lattice spacing: $d_{2\text{-}20(MgO)}=0.1489$nm, $d_{1\text{-}12(MgB2)}=0.1470$nm, $d_{220(MgO)}=0.1489$nm, and $d_{110(MgB2)}=0.1543$nm,; (b) Superimposed diffraction patterns of $(\text{-}111)^*_{MgO}$ and $(2\text{-}21)^*_{MgB2}$ with configuration of the Mg and B layers projected along the $[\text{-}111]_{MgO}$ and $[2\text{-}21]_{MgB2}$ directions. Lattice spacing: $d_{2\text{-}24(MgO)}=0.0860$nm, $d_{\text{-}114(MgB2)}=0.0836$nm, $d_{220(MgO)}=0.1489$nm, and $d_{110(MgB2)}=0.1543$nm.

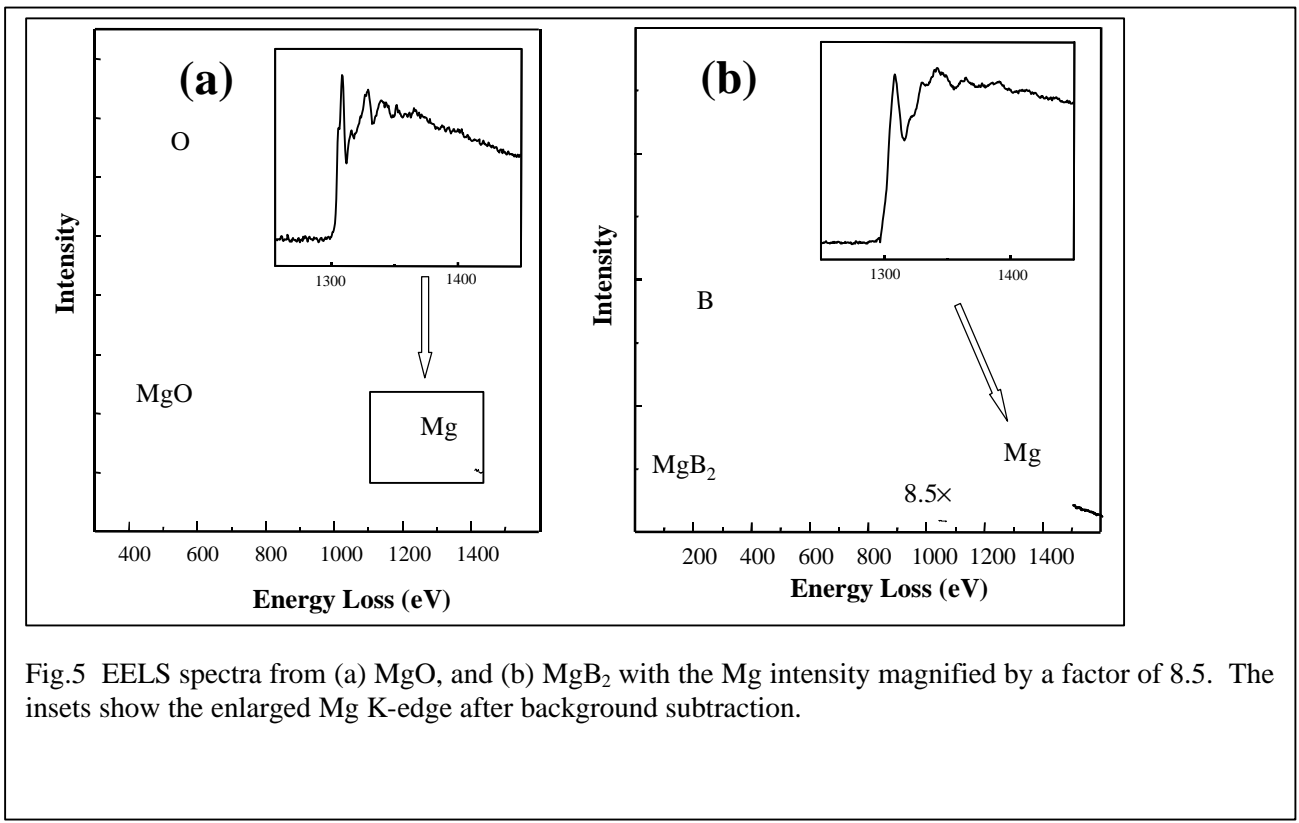

Fig.5 EELS spectra from (a) MgO, and (b) $MgB_2$ with the Mg intensity magnified by a factor of 8.5. The insets show the enlarged Mg K-edge after background subtraction.

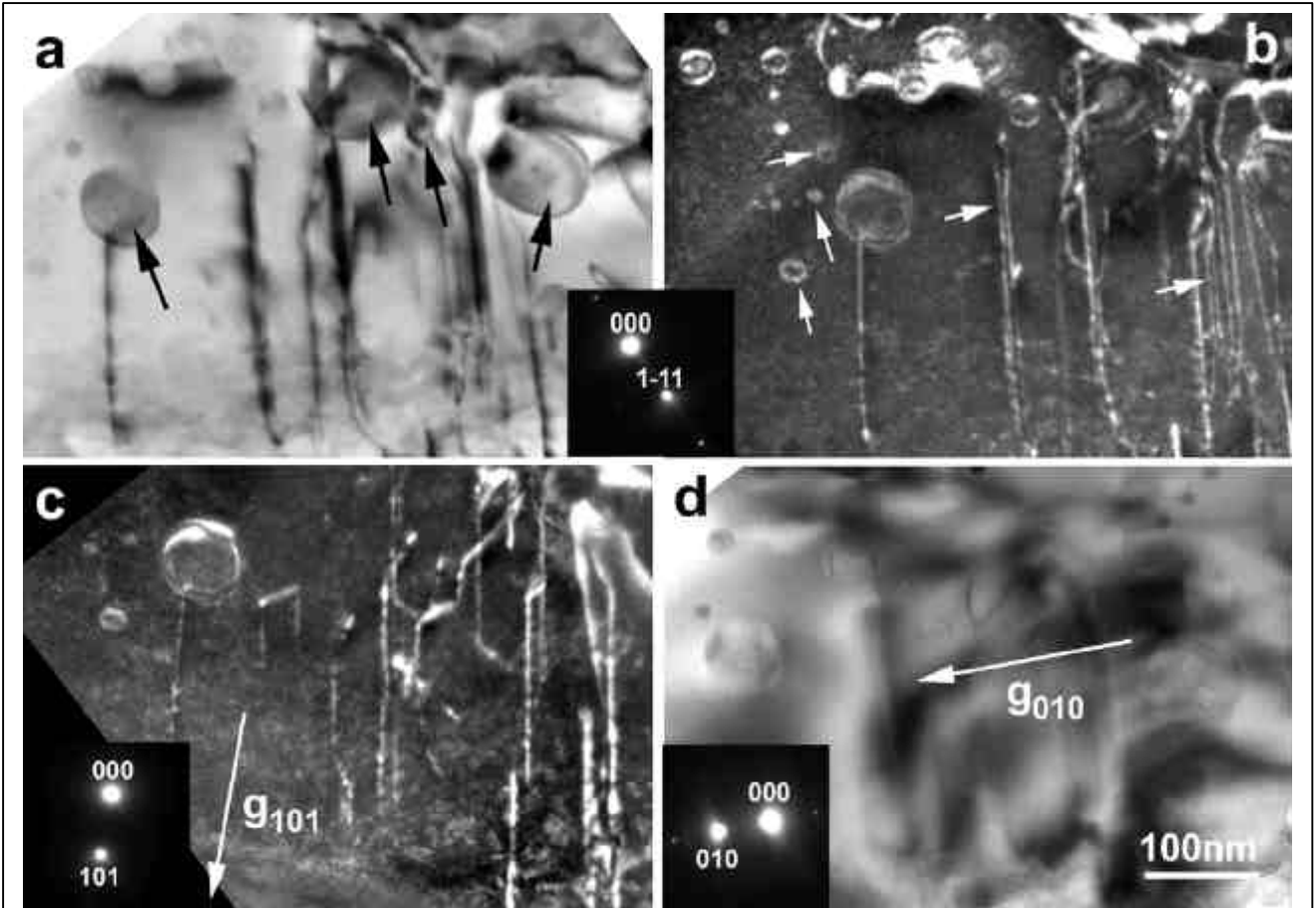

Fig.6 Structural defects in MgB$_2$. Images from the same area using different operating reflections. (a) Bright-field image with **g**=1-11 reflection, (b-d) dark-field image using **g**=1-11 reflection (b), **g**=101 reflection (c) and **g**=010 reflection (d). Note, most of the defects are out of contrast in (d). Large arrows indicate MgO particles and small arrows indicate dislocation loops. Partial dislocations are also visible.

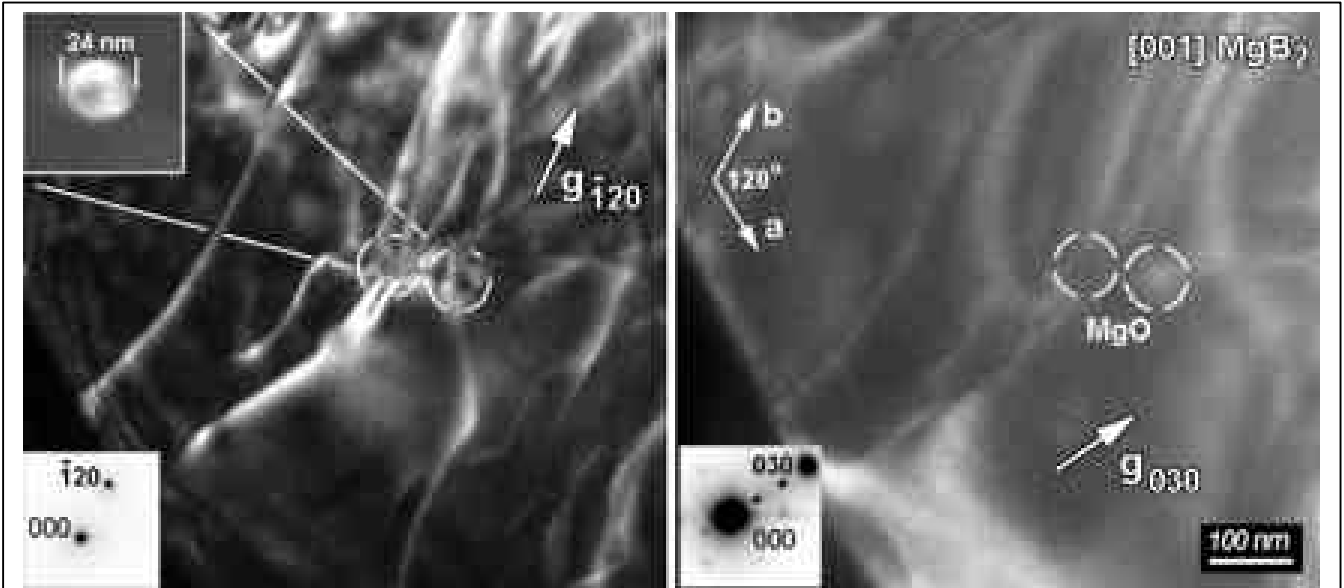

Fig.7 Dark-field images, viewed near along the c-axis of $MgB_2$ crystal with MgO particles as a nucleation center for dislocations. (a) with **g**= $-120_{MgB2}$ reflection. There are two adjacent particles in the area, one imaged using the **g**= $400_{MgO}$ reflection is shown in inset. (b) with **g**=$030_{MgB2}$ reflection. Dislocations are out of contrast, while the particles are still visible.

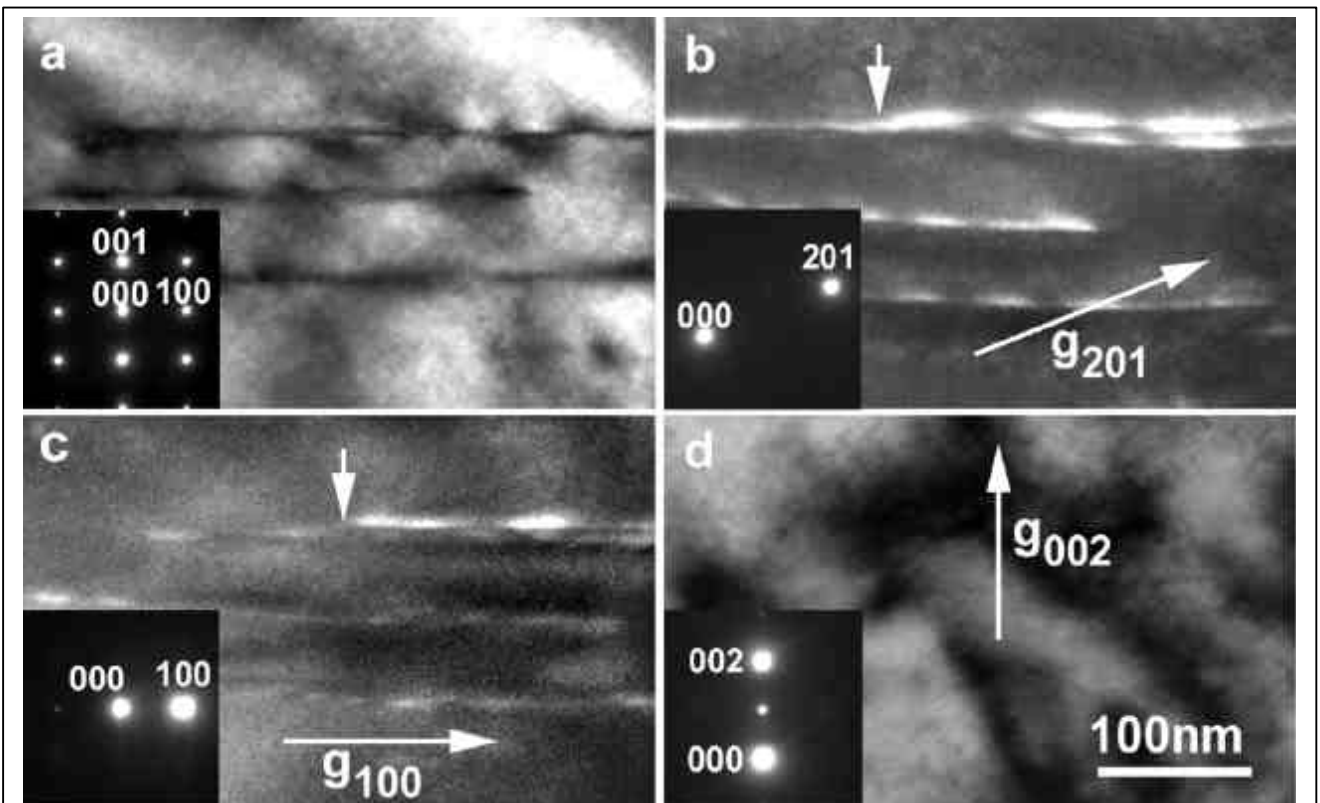

Fig.8 A partial dislocation associated with a stacking fault (marked by an arrow) viewed with the basal plane near edge-on. All the corresponding diffraction patterns are shown in insets. (a) The basal planes are viewed exactly edge-on. (b-d) the basal plane was tilt away ~10° from the zone axis. The dissociation of a perfect dislocation into two partials is clearly visible with the 201 reflection (b) and the 100 reflection (c), but invisible with the 002 reflection.

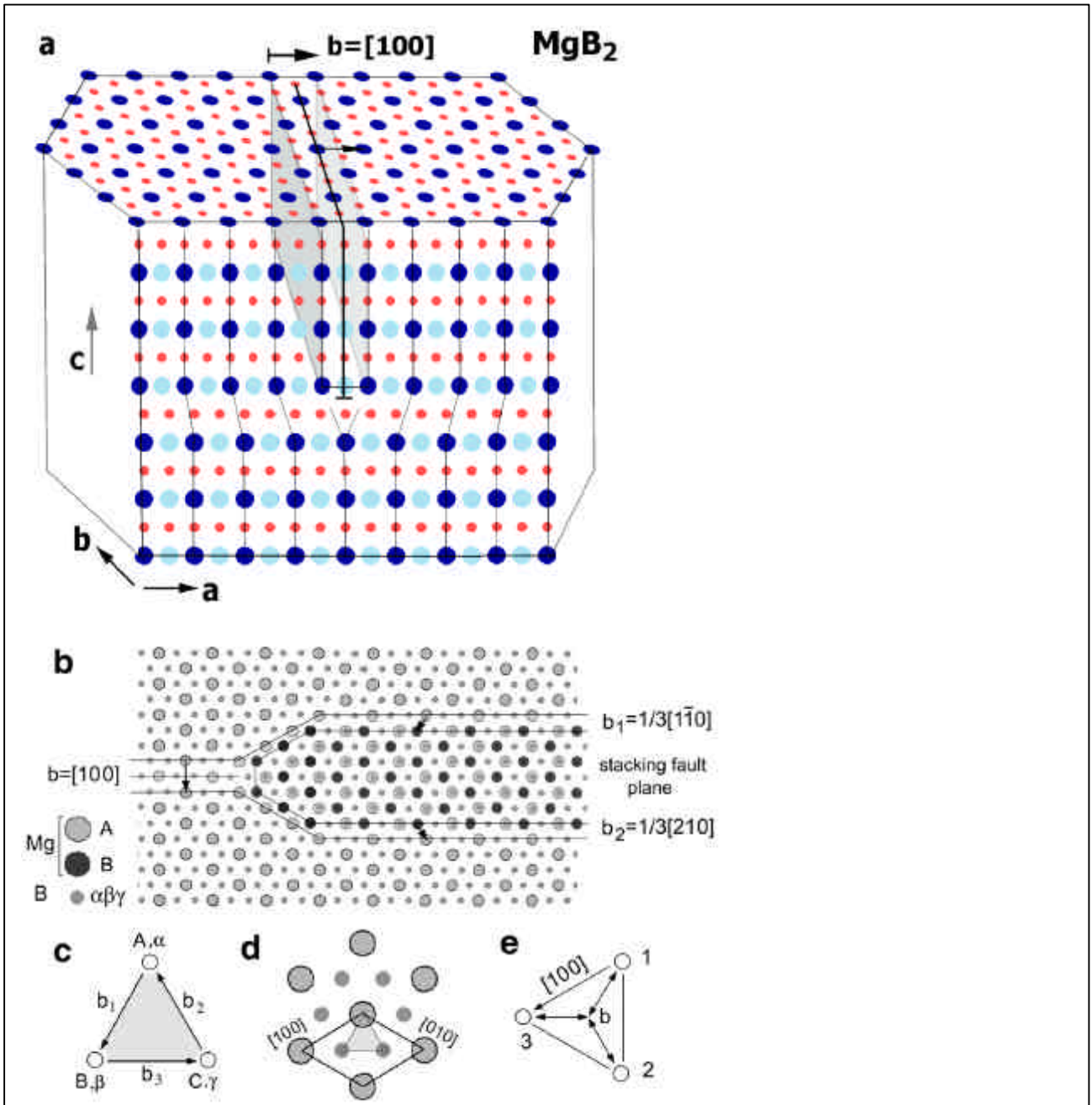

Fig.9(a) Structural model for a perfect edge dislocation with a Burgers vector **b**=[100]. The small and big dots represents B and Mg atoms, while the dark and gray color of the big dots represent the Mg at different depths.  (b)  Structural model for the dissociation of a perfect dislocation into two partials: **b**=[100]=**b₁**+**b₂**=1/3[1-10]+1/3[210].  The stacking sequence across the fault is: …A(βγ)A(βγ)A(βγ) | B(γα)B(γα) | A(βγ)A(βγ)A(βγ)…  where A and B denote the magnesium sublattice and αβγ the boron sublattice, as depicted in (c-e).  For details, see text.

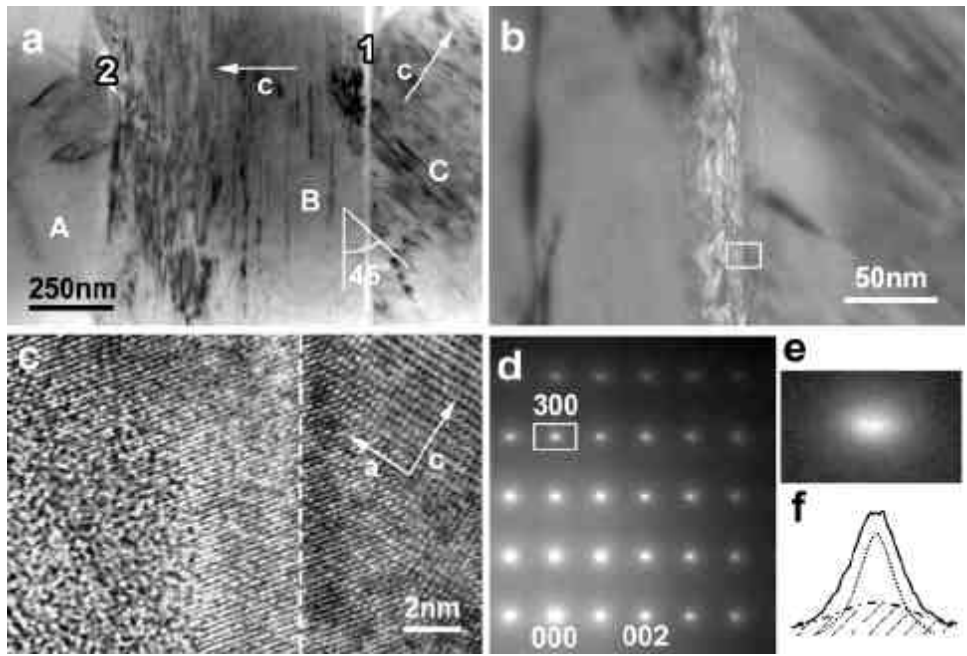

Fig.10 (a) Three grains with two typical grain boundaries in $MgB_2$. "1": amorphous boundary, and "2": structurally intact boundary. The dense parallel line contrast across the amorphous boundary corresponds to the projected basal planes. (b) Enlargement of the amorphous boundary. (c) HREM of enlarged box area in (b) showing local crystalline region in the amorphous layer. (d) Diffraction pattern covering an area similar to the grain B in (a) showing a broadening of the Bragg reflections in the direction perpendicular to the basal planes. (e) Enlargement of reflection of 300 and (f) its intensity profile fitted with the Bragg intensity and diffuse scattering (shaded area).

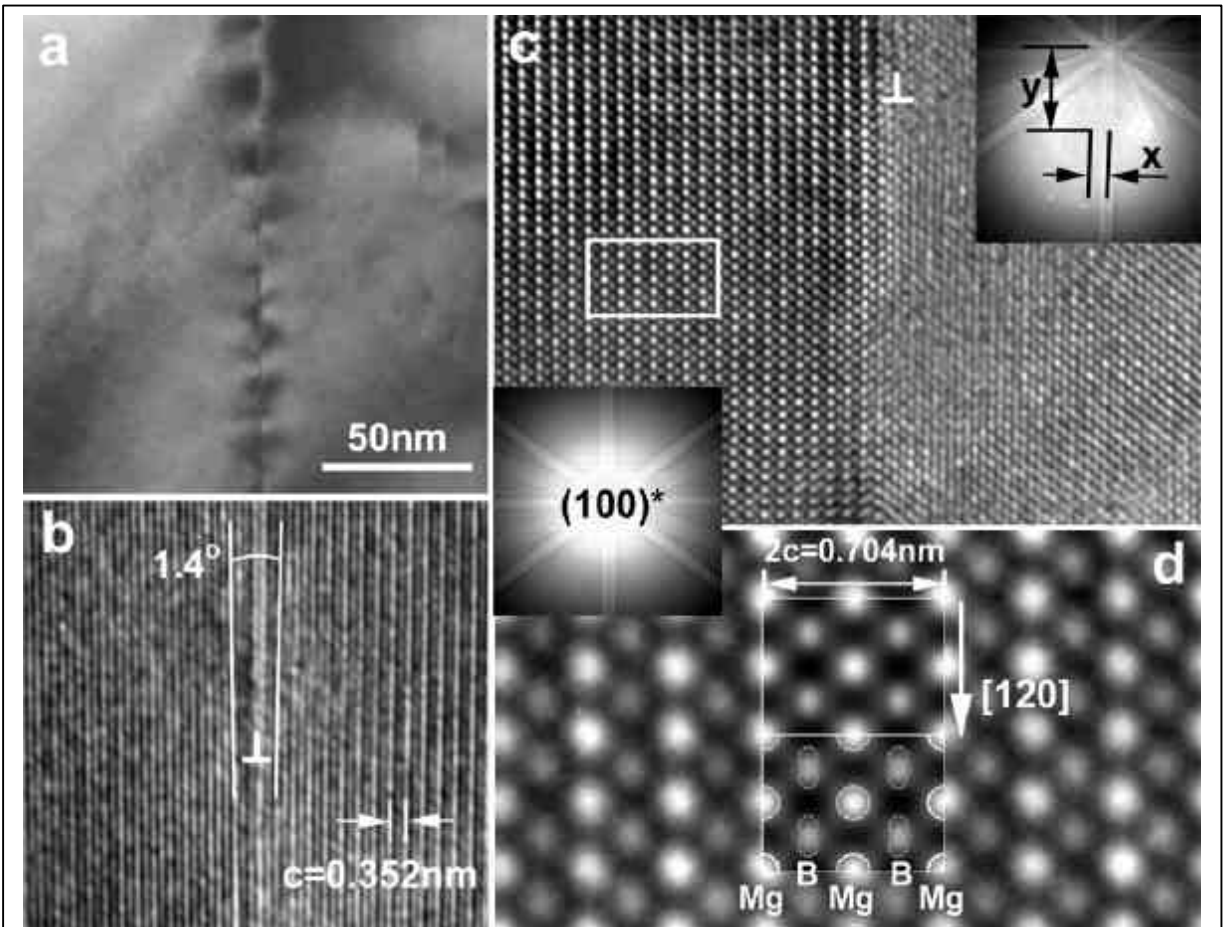

Fig.11 A 4.1° small-angle (001) grain boundary in $MgB_2$. (a) Diffraction contrast associated with interfacial dislocations. (b) Both crystals were tilted away from any zone axes, showing clearly the alternating Mg-B layers (the (002) lattice fringes) and the 1.4° tilt component. (c) The left crystal was tilted to the [100] direction, so both Mg and B atoms are clearly visible. The Kikuchi patterns acquired from both crystals are also included to show the exact misorientation of the grain boundary (x=0.8°, y=3.7°). (d) Enlarged box area of (c) showing the arrangement of B and Mg atoms. The embedded image is the calculated one for defocus value $\Delta f=-60$nm and thickness t=3nm, where the Mg atoms are seen as bright dots, while the B atoms are weak white dots.

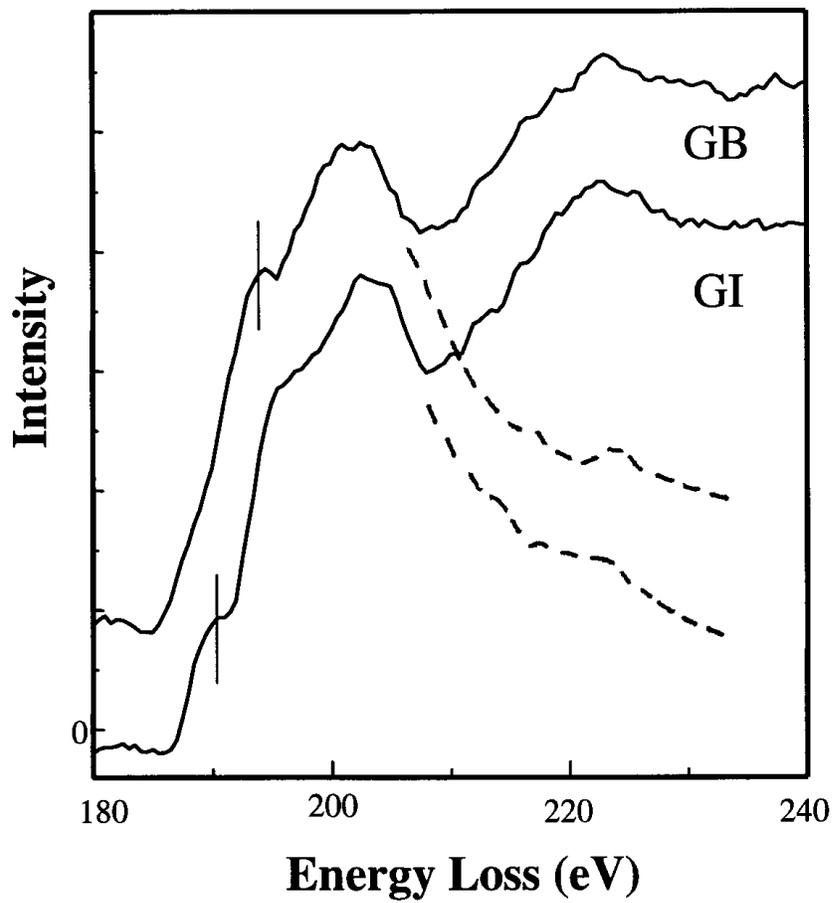

Fig.12 Nano-probe EELS showing the fine structure of the boron K-edge acquired from a grain boundary (GB) and grain interior (GI, 50nm away from the GB). Note the change of the core-loss intensity of the pre-peak marked by the vertical thin lines in the spectra. The dashed lines represent the spectra after plural scattering removal.

20